# IoT-Based Water Quality Monitoring System in Philippine Off-Grid Communities

Jenny Vi Abrajano
*Department of Information Technology*
*Central Luzon State University*
Science City of Munoz, Philippines
abrajano.jenny@clsu2.edu.ph

Khavee Agustus Botangen
*Cyber-physical Systems and IoT Lab*
*Department of Information Technology*
*Central Luzon State University*
Science City of Munoz, Philippines
*Corresponding author:
kwbotangen@clsu.edu.ph

Jovith Nabua
*Department of Information Technology*
*Central Luzon State University*
Science City of Munoz, Philippines
nabua.jovith@clsu2.edu.ph

Jenalyn Apanay
*Department of Information Technology*
*Central Luzon State University*
Science City of Munoz, Philippines
apanay.jenalyn@clsu2.edu.ph

Chezalea Fay Peña
*Department of Information Technology*
*Central Luzon State University*
Science City of Munoz, Philippines
chezaleafay@clsu.edu.ph

*Abstract*—Contaminated and polluted water poses significant threats to human health, necessitating vigilant monitoring of water sources for potential contamination. This paper introduces a low-cost Internet of Things (IoT)-based water quality monitoring system designed to address water quality challenges in rural communities, as demonstrated through a case study conducted in the Philippines. The system consists of two core components. The hardware component of the system, built on Arduino technology and featuring real-time data transmission, focuses on monitoring pH levels, turbidity, and temperature via sensors. The system is equipped to transmit data to a cloud database and send informative messages to mobile numbers, updating users on the status of water supplies. The application component acts as a user interface for accessing and managing data collected by the sensors. The successful deployment of this Water Quality Monitoring (WQM) system not only helps community leaders and health workers monitor water sources but also underscores its potential to empower communities in safeguarding their water sources, thereby contributing to the advancement of clean and safe water access.

## I. Introduction

Water is a fundamental human need necessary for drinking and supporting sanitation and hygiene, sustaining life and health. It is not only an essential resource for human survival but also plays a sanitary, social, and cultural role at the heart of human societies [1]. Access to clean water is a human right and one of the targets in the UN sustainable development goals (SDG 6). Data from the World Health Organization in 2017 shows that around 2.2 billion of the global population still lack basic water service that is available when needed and free from contamination, predominantly in the rural areas of developing nations [2]. Safely managed drinking water services are significantly less accessible in rural areas.

Groundwater constitutes approximately 99% of all liquid freshwater on Earth [3]. In the Philippines, 85% of the piped water supply originates from groundwater [4]. However, only 48% of the 113 million Philippine population receive safely managed or piped water services [5]. Access to safe water poses a significant challenge for a considerable portion of the population. Local government units (LGUs) and water districts (WDs) bear the responsibility of providing safe water services to communities, grappling with issues such as limited funds and capacity. Nevertheless, groundwater remains the sole viable and economical means to extend basic water access to underserved rural populations. Communal wells and hand-operated water pumps are established, becoming the primary sources of drinking water in rural areas. These wells and pumps also serve as the main water sources for vulnerable groups not formally connected or lacking access to public sources.

However, as these water sources are often untreated, they are at risk of contamination and pollution resulting from both anthropogenic and geogenic processes [3]. In rural areas, water source pollution commonly stems from various human activities, including the disposal of organic waste, animal husbandry waste, farm fertilizers, and pesticides [6, 7]. Moreover, the Philippines is susceptible to natural calamities such as typhoons, volcanic eruptions, earthquakes, rain-induced flooding, and landslides which affect the safety of domestic water sources. Natural occurrences can also contribute to groundwater pollution. Seawater intrusion in events of sea level rise may compromise groundwater sanitation. Heavy rains, particularly in areas with inadequate sanitation, can lead to flushing of fecal microbial pathogens and chemicals to the water table [3]. Contamination of domestic water supply sources stands as a primary cause of waterborne diseases, including diarrhea. Unsafe water has severe implications to human health [7]. For instance, in 2017, diarrhea ranked 7th among the ten leading causes of morbidity in the Philippines [8]. Hence, there is a pressing need to monitor and control water quality in these sources, given the potential health risks.

Water quality monitoring (WQM) in rural communities in the Philippines currently relies on the traditional approach of water sampling and analysis. In this method, an individual takes water samples from a source and transports them to a laboratory for subsequent analysis. However, with the advent of wireless sensor networks (WSN), the prospect of smart and real-time WQM has become appealing as devices and communication techniques have advanced. The use of WSNs for WQM has garnered interest due to the low cost of sensor nodes, the capability to collect data at multiple sampling points, and the ability to communicate data to decision makers in a timely manner using low-power communication techniques.

This paper introduces an Arduino-based water quality testing and monitoring system focusing on three water quality parameters: pH, turbidity, and temperature. The system consists of two core components: the first employs sensors to capture real-time values for the three water quality parameters and utilizes a Wi-Fi module to transmit data to the Internet. This component also integrates a GSM module to handle SMS messaging. The second component is a Web application, functioning as the user interface for remote viewing and management of the collected sensor data. The application component includes an Android app version designed for monitoring the test data via smartphones. Our system design departs from typical implementations tailored for urban public water sources. Our goal is to provide a monitoring tool for assessing potential contamination in drinking water supply sources in the most vulnerable populations and off-grid communities in the Philippines. Leveraging solar power and targeting these vulnerable demographics, our innovative solution seeks to empower communities in safeguarding their water sources, thereby improving access to clean and safe drinking water.

## II. LITERATURE REVIEW

Recently, the surge in Internet-of-Things (IoT) advancements has spurred the development of numerous IoT-based solutions for water monitoring. Noteworthy among these are commercial systems like Hach[1] and MetriNet[2], renowned for their reliability and comprehensive functionality. The success story presented by the Public Utilities Board of Singapore [9] showcases the effective implementation of commercial smart-water monitoring technologies in managing critical operational aspects of a water distribution network. These aspects include asset management, leak management, water quality monitoring, meter reading, and conservation. Despite their effectiveness, the widespread adoption of such commercial systems is hindered by their high cost and operational complexity, limiting usage primarily to developed countries and affluent communities [10]. Consequently, there has been a plethora of studies aiming to devise economical yet reliable IoT-based smart solutions to water monitoring. These endeavors seek to leverage existing communication infrastructure to support the implementation of smart applications, thus addressing the needs of vast rural or economically disadvantaged communities.

Various review articles have presented multiple approaches in the realm of smart water monitoring. Notably, Jan et al. [10] offers insights into works on domestic water monitoring. Additionally, Adu-manu et al. [11] and Olatinwo and Joubert [12] published an exclusive review on water quality monitoring based on wireless sensor network (WSN) technologies. A recent survey by Zulkifli et al. [13] encompasses IoT-based water monitoring systems, listing works that integrates recent advances in IoT technology for water quality management.

While the literature provides an extensive array of research on smart water systems, we focus on those closely related to our work, which centers around water quality monitoring (WQM) system utilizing IoT technologies. Pasika and Gandla [14] proposed a low-cost WQM system centered on sensing pH level and turbidity. The system also monitors the water level in the tank, as well as the temperature and humidity of the surrounding atmosphere using a DHT-11 sensor. The sensors are first read by an Arduino Mega2560, then sends the data to NodeMCU, which subsequently uploads the data to the ThingSpeak IoT-Platform. Processed data in ThingSpeak can be accessed through smartphones. Similarly, Almojela et al. [15] proposed a WQM system based on pH, temperature, turbidity, and electrical conductivity. This system also employed an Arduino Nano microcontroller and a NodeMCU to send data to ThingSpeak, with a focus on monitoring drinking water supplied in urban areas.

Geetha and Gouthami [16] proposed a real-time smart WQM system based on parameters pH, temperature, turbidity, and electrical conductivity, specifically designed for in-pipe domestic water. The system also monitors water level in the tank, with decision-making based on preset threshold values conforming to WHO standards. This work utilizes the TI CC3200 microcontroller which has a built-in WiFi module. Data is transmitted to the Ubidots IoT-platform, where the sensor data is compared with preset threshold values. Konde and Deosarkar [17] introduced a smart WQM system employing Field Programmable Gate Array (FPGA) board and aZigbee-based wireless communication module. The system measures water quality parameters such as turbidity, pH, temperature and humidity, and carbon dioxide ($CO_2$), along with monitoring water levels. The work is intended for monitoring large water bodies and resources as part of managing environmental and ecological balance. Similarly, Demetillo et al. [18] presented a comparable work designed for the remote monitoring of bodies of water. This system collects temperature, pH, and dissolved oxygen (DO) as WQM parameters, utilizing an Arduino Mega 2560 microcontroller and a Zigbee transceiver. Moreover, Hong et al. [19] presented a simpler Arduino-based WQM intended for streams and rivers. This system collects parameters such as temperature, pH, turbidity, and total dissolved solids (TDS). Unlike the previous systems, it employed the Arduino Uno R3 microcontroller directly connected to a laptop for data processing and analysis.

A consistent pattern is observed across the various studies in WQM systems. The core controller integrates the variety of sensors, gets data from the sensors, processes the data, and subsequently sends data to wireless communication modules. The wireless communication module uploads data to cloud-based IoT platforms where users are able to access data through web-based applications.

## III. METHODOLOGY

In this section, we present the materials and system design used in this work.

### A. Sensors and Hardware Modules

We connect the temperature, pH, and turbidity sensors to the microcontroller through their respective interface circuits. In the selection of sensors and hardware modules, we opted for open-source Arduino compatible sensors and modules due to their cost effectiveness, satisfactory accuracy levels, and general compliance with standards, as highlighted in the findings of Pereira and Ramos [20]. Moreover, the widespread utilization of Arduino-based sensors in scientific projects has been well-documented [18]. The following are the major components of WQM system:

- **Arduino UNO R3 Development Board ATmega328P CH340** microcontroller board – we opted to use UNO as it is the most used and documented Arduino microcontroller board.
- **16x2 LCD Display –** a 16 characters x 2 lines that is used to display different system data and statuses
- **ESP8266 ESP-01 WiFi Module –** connects the system to WiFi and sends data to Google Firebase.
- **SIM900 GSM GPRS Shield -** sends SMS to concerned local government units.
- **Dallas DS18B20 Waterproof Temperature Sensor -** a single cable with stainless steel sheath sensor which derives its power from its data line itself.
- **Liquid PH Value Detection Sensor -** a 5±0.2V, 5-10mA, and with PH0-14 detection concentration range Arduino compatible sensor.
- **5V DC Turbidity Sensor –** a 40mA (max) sensor with either 0-4.5V analog output or high/low level signal digital output, which operates based on the concentration of total suspended solids in water.
- **32 GB Micro SD card –** used to locally store data collected by the sensors.
- **Piezo Buzzer-** used as an audio-signaling device when testing system parameters.
- **DS3231 RTC module –** a real-time clock module for setting time and alarm.
- **Seeed Studio Solar Panel –** a 0.5W, 70x55x3 mm, 100 mA solar panel compatible with the microcontroller.
- **12V 1.3Ah Pilum Rechargeable Battery –**used to power the microcontroller board in the absence of sunlight.

---

[1] https://www.hach.com/
[2] https://www.analyticaltechnology.com/us/water-monitoring/metrinet/

We also use a water and dust proof 36x20x11cm enclosure for integrating the system modules and to organize the connecting wires.

*B. System Design*

The top-level architecture of the WQM system is illustrated in Fig. 1. The hardware module collects data via the sensors, and sends the collected data to Google Firebase from which the system application derives the information being provided to users. The Google Firebase data log (i.e., in JSON format) has a "readings" node with child nodes, each child node representing a single reading captured by a sensor. A reading has attributes such as timestamp, sensor ID, parameter measured, value, unit, status, and location (e.g., *readings: { reading_id_1: { timestamp: "2023-01-15T08:00:00Z", sensor_id: "sensor_1", parameter: "pH", value: 7.2, unit: "pH", status: "Normal", location: "Loc 1 - Community A"},…}*). There is also an "alerts" node with child nodes, each child node representing an alert triggered based on predefined thresholds. Alerts have similar attributes to readings but also include a message describing the alert condition (e.g., *alerts: { alert_id_1: { timestamp: "2023-01-15T08:15:00Z", sensor_id: "sensor_1", parameter: "pH", value: 9.2, unit: "pH", status: "High", location: "Loc 2 - Community A", message: "pH level above normal"},…}*). The database can be queried to retrieve readings, track trends, and monitor for abnormal conditions in real-time.

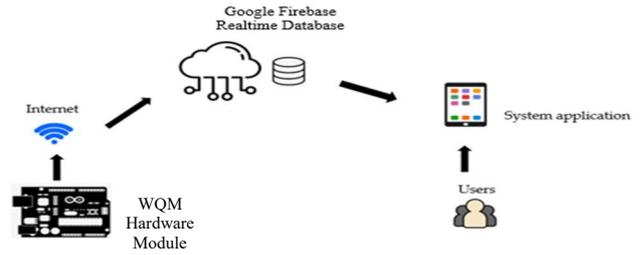

Fig. 1. WQM system architecture.

Fig. 2 illustrates the integration diagram of the components comprising the WQM system. Our material selection aligns with our objective to create a system that is cost-effective, portable, energy-efficient, IoT-enabled, and real-time. Similar criteria were targeted in the survey of works conducted by Jan et al. [10]. The system is designed to collect data on water temperature, pH level, and turbidity, as these parameters are commonly considered in most works assessing domestic water quality [10, 11]. While the World Health Organization (WHO) provides an extensive list of water quality parameters, water authorities typically focus on a limited set, which includes these three parameters.

Jan et al. [10] has enumerated various studies demonstrating inter-correlation among WQM parameters, suggesting that certain parameters can be inferred from the measured values of others. Due to this intercorrelation, employing a limited set of parameters may suffice for determining water quality. Among the most recommended parameters in most IoT-WQM studies [10, 16, 11, 13] are turbidity, temperature, pH, oxidation-reduction potential, and electrical conductivity. In our study, we excluded oxidation-reduction potential since this parameter correlates with pH and temperature [10]. Similarly, electrical conductivity is correlated with turbidity and pH [21, 10].

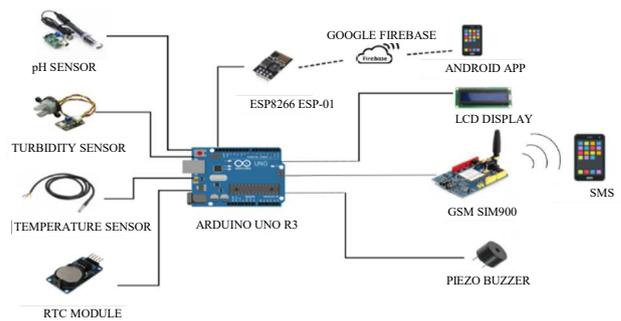

Fig. 2. WQM system components.

To conduct a test, we perform 5 successive reads for each sensor with a 30-seconds interval and then check the average value against the drinking water parameter thresholds. The thresholds we adopt are in accordance with the guidelines established by the World Health Organization (WHO), the United States Environmental Protection Agency, and the Philippine National Standards for Drinking Water [10, 22, 23], as detailed in Table 1.

TABLE I. PARAMETER RANGE OR CONDITION FOR DRINKING WATER

| Parameter | Threshold |
|---|---|
| Turbidity | <5 NTU |
| Temperature | 15 - 40 Degrees Celsius |
| pH | 6.5 – 8.5 pH |

For each parameter, the buzzer activates when the computed value falls below or exceeds the designated threshold. During each test, the sensor data and computed data are stored in the system's SD card memory module as log information, inclusive of timestamps. These data can later be retrieved from the SD card in CSV format. In instances when internet connectivity is available, the test data is transmitted as a CSV file to the cloud database. Simultaneously, an SMS is sent to community health authorities responsible for monitoring domestic water sources, alerting them to the situation and prompting appropriate actions. Anticipating deployment in rural villages where a stable source of electricity is uncertain, we have designed the system to be solar panel-ready, capable of powering the water quality monitoring (WQM) components. Additionally, the system includes a battery backup to ensure functionality during periods of low sunlight or no sunlight at all.

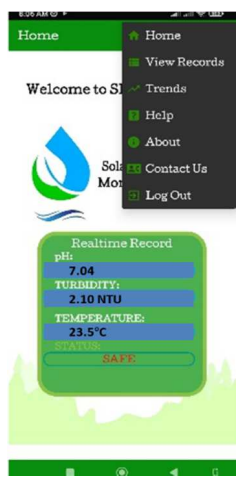

Fig. 3. The Web application component of the system.

Fig. 3 shows the web application that has been developed to provide access to logs stored in the Google Firebase database and generate summary reports of conducted tests. This application is designed to streamline water monitoring tasks for community leaders, health workers, and concerned authorities. It serves as a valuable tool in their decision-making processes for developing effective approaches to improve water service provisions.

## IV. RESULTS AND DISCUSSION

### A. Deployment of the WQM System

Fig. 4 displays images of the WQM system installed for use in a communal water pump within a rural community of indigenous peoples in the mountainous portion of Central Luzon, Philippines. Situated 15 kms away from the main community center and lacking public water system, this community relies on a hand-operated water pump – a common equipment used to source water in the rural Philippine communities without access to public water systems. Given the outdoor setting, we constructed a two-layered steel housing measuring 108cm in height and 60x60cm in dimensions, featuring a galvanized roof to accommodate the hardware components of the WQM system. The solar panel is integrated into the roof, while the controller enclosure and battery are placed on each layer, respectively. The total cost of implementing the system, including hardware, housing, and fabrication, amounts to $300 USD.

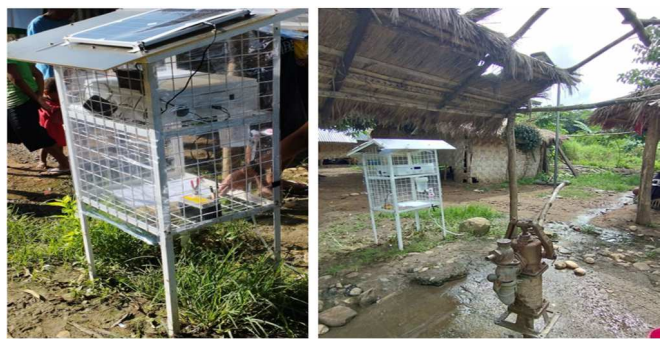

Fig. 4. System deployment in a communal hand-operated water pump.

It is essential to highlight that the steel housing enclosing the Water Quality Monitoring (WQM) system components is securely locked and can only be opened and operated by the community leader or an individual authorized by the barangay with jurisdiction over the community. In the Philippines, a barangay represents the smallest political unit, and each municipality or city is composed of multiple barangays. Therefore, a test can only be initiated in the presence of the individual holding the key to the steel housing.

The system is activated by pressing a designated power button. Once the controller and sensors are operational, the LCD screen displays the message "Ready," indicating that the system is prepared for use. To conduct a test, water samples from the pump must be poured to a specified level or higher into the water bucket containing the sensors. The bucket is designed with a drain cap to facilitate quick removal and replacement of water samples when necessary. Each test cycle consists of five readings. During each cycle, the water level is initially tested, ensuring it reaches the indicated level or higher in the bucket. Subsequently, the three parameters—pH, turbidity, and temperature—are tested sequentially. The results of each sensor reading are displayed on the screen, and the average value from the five readings is computed for each parameter. The system evaluates the average value for each parameter against the predefined thresholds. When the average value falls below or exceeds the threshold, the buzzer activates, accompanied by the display of the computed value and a descriptive result (e.g., "pH PASSED – Above Limit" or "pH FAILED – Below Limit").

Based on the test results, the community fetching water from the pump is promptly informed through a predetermined indicator agreed upon by the community members. A blue flag is raised at the water source, signaling that the water may be deemed safe to drink. Conversely, a red flag is raised to indicate that the water may not be safe for consumption. In instances of red flags, the community is generally advised to boil the water before drinking. Utilizing the SMS capability of the system, the test results are concurrently forwarded to the phone numbers of the barangay captain, the barangay council member that is assigned to the community, and the barangay health officer. This ensures that timely and relevant actions can be undertaken, such as disseminating information on necessary community actions, providing water treatment services, or arranging for the delivery of water supplies.

The operationalization of the system in this deployment setting can follow a predefined schedule, such as once a week, or can be triggered by observable changes in the water properties. For instance, during the rainy season, the community often experiences flooding and landslides, impacting the water supply, particularly the turbidity parameter. Turbidity, which refers to the cloudiness of water due to suspended particles like mud, sand, chemicals, and organic elements, can indicate pollution, the presence of pathogens, and may affect the water's appearance and taste negatively [22]. Testing the water source after heavy rainfall becomes crucial during this period. In contrast, during summer seasons, observable changes in water properties are typically less pronounced. Nevertheless, regular testing remains essential due to higher occurrences of microorganism growth at elevated temperatures [24]. The increase in water temperature can trigger chemical reactions that may impact the potability of water [22].

We have implemented adjustments in how the system interprets threshold values, particularly for the temperature parameter. Unlike other parameters, we do not consider the threshold values for temperature as strict determinants for the safety of the water source. Even though the lower limit specified in Table 1 is 15 degrees Celsius, the system does not categorize results below 15 degrees as a "FAILED" test. This decision is informed by the cold weather conditions often experienced in mountainous areas, which can significantly lower the temperature of their water sources. While water temperature can influence the chemical, biological, and physical properties of water, as well as potential health effects, low water temperature does not have a significant impact on potability compared to elevated temperatures. Cool water is generally more palatable than warm water, and temperature primarily affects the acceptability of various inorganic constituents and chemical contaminants that may influence taste. In contrast, high water temperature can enhance the growth of microorganisms and potentially lead to taste, odor, color, and corrosion issues [24]. It is worth noting that, according to the WHO report [25] on a survey of 104 countries, 18 countries incorporate a regulatory value for temperature in water quality. The review further specifies that none of the values for temperature were mandatory; instead, they were used as guiding levels or operational goals.

TABLE II. SENSOR CALIBRATION

|   | Percentage Differences on Seven (7) Samples | Mean Percentage Difference |
|---|---|---|
| pH | (2.16, 2.50, 1.55, 2.55, 2.04, 2.82, 1.34) | 2.14% |
| Turbidity | (5.35, 4.65, 2.35, 6.74, 4.44, 7.79, 6.19) | 5.36% |
| Temperature | (4.05, 4.41, 4.96, 5.22, 2.78, 3.51, 3.39) | 4.05% |

*B. System Evaluation*

We assess the calibration of our sensors by comparing the observed data values with those obtained from testing samples provided by a commercial water refilling station. On seven separate occasions, we brought water samples to the refilling station and tested them using our approach. Subsequently, immediately after conducting the test with our Water Quality Monitoring (WQM) system, we subjected the samples to testing at the water refilling station. Table II presents the percentage difference between the refilling station and WQM test results for the three parameters. The relatively small variation in the pH values suggests good agreement between the two methods of measuring pH. However, for turbidity and temperature, there is a relatively larger percentage of variation among the values. We need to seek an additional water testing facility to compare the results, and if a similar variation is observed, then we may need to consider replacing the respective sensors.

We invited 10 IT experts specializing in Networks and Software Development to evaluate the Water Quality Monitoring (WQM) system using criteria adopted from ISO/IEC 25010 for assessing systems and software quality requirements. Both the hardware and application components of the system are assessed on a rating scale from 1 (lowest) to 5 (highest). Fig. 5 displays the consolidated result of the evaluation.

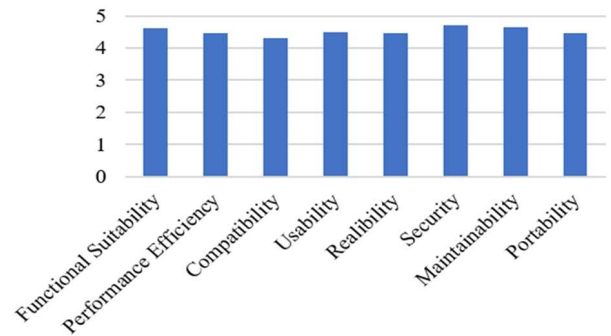

Fig. 5. WQM system evaluation by IT experts.

There is a relatively high evaluation rating in all the criteria; however, the experts identified compatibility, efficiency, reliability, and portability as areas for improvement. Suggestions were made to enhance the compatibility of the hardware module by allowing the addition of sensors that can be plugged in and configured through the application component. Concerns regarding efficiency and reliability include potential degradation of sensors leading to inaccurate readings and challenges in transmitting data to the application database when there is no internet, which is particularly relevant in rural communities lacking assured internet connectivity.

Additionally, there are environmental resilience concerns, given the system's deployment in an outdoor environment with constant exposure to changing elements such as heat and rain, which may affect the wear and tear properties of the system components. Regarding portability, experts recommended a miniaturized version of the hardware component for ease of use and deployment. The system enclosure should be designed to be compact, lightweight, durable, and water-resistant.

## V. Conclusion and Future Works

This paper introduces an innovative Arduino-based Water Quality Monitoring (WQM) system designed for deployment in rural communities, particularly those facing challenges in accessing safe drinking water. With a focus on assessing three crucial parameters—pH, turbidity, and temperature—the system leverages IoT technologies to enable real-time monitoring and reporting. The deployment of the WQM system in a rural community has shown promising results, the system was able to provide real-time alerts and updates of water quality parameter to the mobile numbers of concerned community leaders. The successful implementation of the prototype WQM system represents a significant step towards addressing water quality challenges in underserved communities. Since water quality monitoring and expertise is commonly held by specialized institutions. Now communities are empowered to monitor and safeguard their water sources.

Future enhancements include: *i*) **Improved enclosure design** - the development of a more compact, lightweight, durable, and weather-proof system enclosure is crucial for enhancing the portability of the system. *ii*) **Flexible and adaptable configuration** - recognizing the unique water characteristics across different regions in the Philippines, future iterations will focus on creating a flexible and easy-to-configure WQM system. This adaptability is essential to accommodate variations in regional water quality standards, ensuring that the system can be easily adjusted to meet the specific needs of different areas. *iii*) **Water quality index (WQI) computation** - an innovative approach involves computing a Water Quality Index (WQI), a single value that provides a comprehensive assessment of water quality. This simplification aims to streamline decision-making processes by condensing multiple parameters into one value. The potential integration of machine learning techniques, as surveyed in previous approaches [10], becomes particularly relevant, especially in scenarios where multiple sensors at various sites perform frequent tests. *iv*) **Enhanced web application dashboard** - an enhanced web application dashboard such as the PhiGo website dashboard [4] will support the real-time monitoring of data from multiple sensors at various sites.